%% file: sample-sigconf-authordraft.tex
\documentclass[manuscript]{acmart}
\AtBeginDocument{%
  }

\setcopyright{acmlicensed}
\copyrightyear{2018}
\acmYear{2018}
\acmDOI{XXXXXXX.XXXXXXX}
\acmConference[Conference acronym 'XX]{Make sure to enter the correct
  conference title from your rights confirmation email}{June 03--05,
  2018}{Woodstock, NY}
\acmISBN{978-1-4503-XXXX-X/2018/06}

\begin{document}

\title{In the Middle, Not on Top: AI-Mediated Communication for Patient–Provider Care Relationships}


\author{Ut Gong}
\authornotemark[1]
\affiliation{%
  \institution{Columbia University}
  \city{New York}
  \country{United States}}
\email{ug2155@columbia.edu}

\author{Yibo Meng}
\authornotemark[1]
\affiliation{%
  \institution{Cornell University}
  \city{New York}
  \country{United States}}
\email{larst@affiliation.org}

\author{Qihan Zhang}
\affiliation{%
  \institution{The Sixth Primary School of Qianxi County}
  \city{Hebei}
  \country{China}}
\email{zhang.qihan@outlook.com}

\author{Xin Chen}
\affiliation{%
  \institution{Universidad Politécnica Pe Madrid}
  \city{Madrid}
  \country{Spain}}
\email{xin.c@alumnos.upm.es}

\author{Yan Guan}
\affiliation{%
  \institution{Tsinghua University}
  \city{Beijing}
  \country{China}}
\email{guany@tsinghua.edu.cn}

\renewcommand{\shortauthors}{Gong et al.}

\begin{abstract}
Relationship-centered care relies on trust and meaningful connection. As AI enters clinical settings, we must ask not just what it can do, but how it should be positioned to support these values. We examine a "middle, not top" approach where AI mediates communication without usurping human judgment. Through studies of CLEAR, an asynchronous messaging system, we show how this configuration addresses real-world constraints like time pressure and uneven health literacy. We find that mediator affordances (e.g., availability, neutrality) redistribute interpretive work and reduce relational friction. Ultimately, we frame AI mediation as relational infrastructure, highlighting critical design tensions around framing power and privacy.
\end{abstract}

\ccsdesc[500]{Human-centered computing~Human-computer interaction (HCI)}
\ccsdesc[300]{Human-centered computing~Empirical studies in HCI}
\ccsdesc[200]{Human-centered computing~Collaborative and social computing}
\ccsdesc[100]{Human-centered computing~Interaction design}

\keywords{
relationship-centered care,
AI-mediated communication,
human--AI interaction,
healthcare communication
}

\begin{teaserfigure}
  \includegraphics[width=\textwidth]{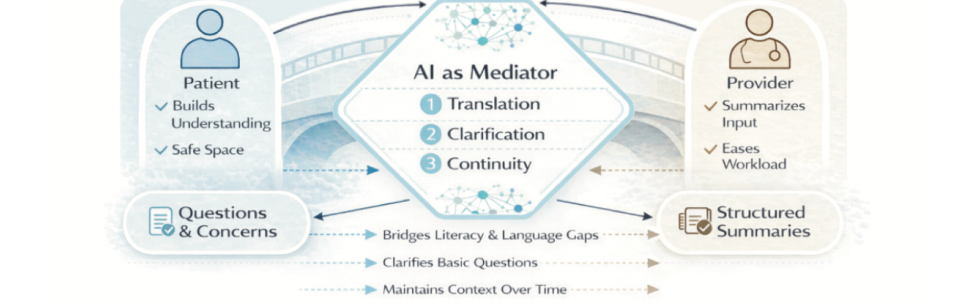}
  \caption{AI functions as relational infrastructure to support translation and continuity, preserving human–human care}
  \label{fig:teaser}
\end{teaserfigure}

\received{20 February 2007}
\received[revised]{12 March 2009}
\received[accepted]{5 June 2009}

\maketitle
\input{text/part1_intro}
\input{text/part2_CLEAR}
\input{text/part3_AIPosition}
\input{text/part4_disussion}
\input{text/part5_conclusion}

\bibliographystyle{ACM-Reference-Format}
\bibliography{sample-base}

\end{document}

%% file: text/part1_intro.tex
\section{Introduction}

Artificial intelligence (AI) is increasingly incorporated into healthcare communication, often framed as a means to improve efficiency, scalability, or personalization \cite{jacobs2021designing, cho2025conversational, 10.1145/3311957.3359433, Schiavo2025-to}. Yet healthcare interaction is fundamentally relational: effective care depends on trust, mutual understanding, and sustained patient--provider relationships \cite{kocielnik2022automated, mandal2025apomediation, kim2025steering}. This tension raises a design question for relationship-centered care: \textbf{how can AI be positioned to support patient--provider relationships without replacing or displacing human interaction?} \cite{jacobs2021designing}.

This question becomes especially pressing in \textit{resource-limited} care contexts---where access to care exists, but structural constraints make mutual understanding difficult to achieve. In such settings, providers may see \textbf{40--80 patients per day} with \textbf{5--7 minute} consultations on average, leaving little opportunity for iterative explanation or clarification. These pressures interact with uneven and sometimes low health literacy, linguistic diversity, and discontinuity of care, producing breakdowns that are not primarily about effort or expertise, but about mismatched communicative capacities between patients who need time to process complex information and providers who lack capacity to support that processing in a compressed encounter \cite{brewer2025designing, liu2025designing, hertzum2025implementing}. At the system level, resource limitation is also measurable (e.g., medically underserved regions defined as having fewer than \textbf{1.5 physicians per 1,000 residents}).

In this paper, we explore one possible positioning---\textbf{AI in the middle, not on top}---where AI mediates communication\cite{hancock2020ai} between patients and providers without assuming clinical authority or speaking on behalf of either party \cite{mandal2025apomediation}. Rather than treating AI as a conversational proxy or decision-maker, this framing conceptualizes AI as relational infrastructure that supports clarification, preparation, and continuity while preserving human judgment and accountability \cite{hertzum2025implementing, yang2024rethinking}.

Our argument is grounded in formative and user studies of CLEAR, an AI-mediated communication system designed for asynchronous patient--provider messaging \textbf{(formative: 6 providers, 20 patients; user: 8 providers, 20 patients)}. CLEAR was developed in response to constraints identified in formative work, including limited clinical time \cite{jacobs2021designing}, uneven medical literacy \cite{brewer2025designing}, linguistic diversity \cite{liu2025designing}, and discontinuity of care providers \cite{hertzum2025implementing}. Across our studies, we observed that positioning AI as a mediator can help patients better understand medical information and articulate concerns over time, while enabling providers to prepare responses more efficiently without ceding authority. These observations suggest that AI's structural affordances---such as availability, consistency, and relative neutrality---can help bridge human--human communication gaps under resource-constrained conditions \cite{kocielnik2022automated, cho2025conversational}.

Building on these findings, we use CLEAR as a grounded case to examine what \textit{AI in the middle} can enable and complicate for patient--provider care relationships. We contribute to workshop discussions by articulating this positioning as one design possibility, reflecting on its relational benefits and limitations, and inviting comparison with alternative ways AI might be situated within relationship-centered care \cite{yang2024rethinking}.

%% file: text/part2_clear.tex
\section{CLEAR: AI-Mediated Communication in Practice}

CLEAR is an AI-mediated communication system for asynchronous messaging between patients and providers in resource-limited care contexts. It targets barriers such as brief consultations, uneven health literacy, and discontinuity of care that make shared understanding hard to achieve. Rather than acting as a proxy clinician, CLEAR positions AI as a mediator that operates \emph{between} patients and providers to support interpretation and communication.

\textbf{Patient-facing support.} CLEAR helps patients revisit care instructions after a visit, translate clinical jargon into more accessible language, and draft follow-up questions over time. As an \emph{asocial} resource, it reduces the social cost of asking ``basic'' or repetitive questions.
\textbf{Provider-facing support.} CLEAR organizes patient messages into actionable context (e.g., summarizing concerns and surfacing items that need clarification), helping providers respond more efficiently, especially when continuity is fragmented.

\textbf{Illustrative vignette.} After a brief visit, a patient remains unsure about a medication change. At home, they use CLEAR to restate instructions in simpler terms and draft questions they hesitated to ask in-clinic. The provider then reviews a compact summary of these uncertainties and addresses them directly, preserving clinician authority while reducing conversational overhead.
\textbf{Observed outcomes.} In our studies, patients reported increased confidence in understanding and communicating about their care, and providers reported reduced preparation burden. Participants generally interpreted CLEAR not as replacing human interaction, but as infrastructure that helps sustain patient--provider communication under real-world constraints.

%% file: text/part3_AIPosition.tex
\section{Findings: CLEAR as a Mediated Relationship Case}

\textbf{F1. Asocial clarification lowered the social cost of not understanding.}
Patients described that asking questions to an AI mediator felt safer than asking a clinician directly---especially when they worried their questions were ``too basic,'' repetitive, or would take too much time. This reduced embarrassment and hesitation, letting patients revisit information and refine questions at their own pace. Participants did not frame this as replacing the clinician; instead, it helped them arrive at the next interaction with clearer uncertainties and more precise questions.

\textbf{F2. Temporal redistribution improved question quality and reduced encounter pressure.}
Participants described a shift in \emph{when} understanding work happened. Rather than forcing comprehension into a short, high-pressure consultation, clarification moved to moments before and after the encounter (e.g., reviewing instructions at home; drafting follow-ups asynchronously). Providers described this as reducing ``in-the-moment'' conversational overhead and making it easier to address the most salient uncertainties when time was limited.

\textbf{F3. Continuity support helped reconstruct narratives when care was discontinuous.}
When patients see multiple providers, both parties face the burden of repeatedly rebuilding clinical context. Mediated messaging and summarization bridge these gaps by maintaining a legible record of past concerns and explanations, preventing misunderstandings from compounding across disjointed visits.

\textbf{F4. Perceived neutrality and consistency shaped trust but also constrained nuance.}
Participants often framed the mediator as more ``consistent'' than hurried human conversation, particularly for translating jargon into accessible language. This consistency supported confidence and reduced confusion. At the same time, consistency risked flattening nuance: a mediator can make information feel definitive even when clinical interpretation is provisional or contested.

\textbf{F5. Mediation introduces relational risks: foregrounding, stabilization, and privacy.}
Systems prioritize specific details, risking "narrative stabilization" where a summary is treated as the authoritative record despite omitting nuance or uncertainty. Furthermore, digital mediation expands the privacy surface area; sensitive disclosures can persist and be shared across time or providers beyond the patient's original intent.

Together, these findings suggest that the value of ``AI in the middle'' is not that it answers medical questions, but that it changes the interactional conditions under which patients and providers reach shared understanding---redistributing interpretive labor across time, reducing social friction, and creating continuity artifacts, while also introducing framing power and privacy risks that must be explicitly designed for.

%% file: text/part4_disussion.tex
\section{Positioning AI in the Middle, Not on Top}

Building on these findings, we argue for a specific positioning of AI in relationship-centered care: \textbf{AI in the middle, not on top}. This framing treats AI not as a proxy clinician or autonomous decision-maker, but as \textbf{relational infrastructure} that mediates communication between patients and providers. The core premise is positional: AI should support the interactional work required for shared understanding---clarifying, preparing, and sustaining continuity---while leaving clinical interpretation and accountability with human providers.

\textbf{What ``in the middle'' entails.}
In this positioning, AI functions as a mediator that (1) helps patients interpret and formulate concerns outside the time pressure of clinical encounters, and (2) helps providers reconstruct context and prepare responses without generating medical decisions on their behalf. This differs from ``AI on top'' roles where AI answers in place of clinicians, delivers recommendations with implied authority, or becomes the primary interface patients rely on for care guidance.

\textbf{Mechanisms of relational mediation.}
Our findings suggest three interactional mechanisms through which AI-in-the-middle can support care relationships under resource constraints:
\begin{enumerate}
    \item \textbf{Temporal redistribution:} shifting interpretive labor before and after the encounter so that short consultations can focus on resolving the most salient uncertainties.
    \item \textbf{Reduced social friction:} providing an asocial channel for clarification that lowers embarrassment and hesitation, enabling patients to ask more questions and refine them over time.
    \item \textbf{Continuity artifacts:} carrying forward legible context across interactions, particularly when care is discontinuous, reducing repeated storytelling and reconstruction burden.
\end{enumerate}
These mechanisms help explain why participants did not experience the mediator as replacing clinicians; instead, it altered the conditions of communication such that human--human interaction could become more effective.

\textbf{Why positioning matters.}
Positioning shapes both expectations and risks. When AI is ``in the middle,'' its outputs are framed as interpretive aids and communicative scaffolds rather than authoritative clinical judgments. This can help preserve provider responsibility while still leveraging AI's availability and consistency. At the same time, mediation necessarily introduces framing power: what the system highlights, summarizes, or rephrases can influence what becomes salient in the relationship. Moreover, by producing persistent records and summaries, AI-in-the-middle can increase privacy exposure and create narrative stabilization effects that may be difficult to undo.

\textbf{Design implication of the stance.}
Treating AI as relational infrastructure implies that the system should be designed to (a) make its mediating role explicit, (b) preserve space for uncertainty and nuance rather than presenting summaries as final, (c) support patient control over what is shared and persisted, and (d) keep clinicians in the loop as the locus of accountability. In the next section, we unpack these tensions and boundary conditions to invite workshop discussion about when ``AI in the middle'' is appropriate and what safeguards it requires.

%% file: text/part5_conclusion.tex
\section{Conclusion}

This workshop paper argues for a relational positioning of healthcare AI: \textbf{AI in the middle, not on top}. Grounded in our studies with CLEAR, we show how AI-mediated communication can redistribute interpretive work across time, reduce social friction around clarification, and support continuity when care is resource-limited while also introducing framing power and privacy risks. We offer this positioning not as a universal solution, but as a design stance and set of open questions for workshop discussion on where AI should sit in relationship-centered care and what safeguards that role requires.